\documentclass[useAMS,usenatbib]{mn2e}
\usepackage{amsmath}
\usepackage{amssymb}
\usepackage{graphicx}
\usepackage{subfig}
\usepackage{epsfig}
\usepackage{caption}
\usepackage{breqn}
\usepackage{changebar}
\usepackage{xcolor}
\usepackage{color,soul}

\title[CEMP stars: possible hosts to carbon planets in the early universe]{CEMP stars: possible hosts to carbon planets in the early universe}
\author[N. Mashian and A. Loeb]{Natalie Mashian$^{1}$\thanks{nmashian@physics.harvard.edu}, Abraham Loeb$^{1}$\thanks{aloeb@cfa.harvard.edu}\\
$^{1}$Harvard-Smithsonian Center for Astrophysics, 60 Garden Street, Cambridge, MA 02138, USA}

\begin{document}

\pagerange{\pageref{firstpage}--\pageref{lastpage}} \pubyear{2015}

\maketitle

\label{firstpage}


\begin{abstract}
We explore the possibility of planet formation in the carbon-rich protoplanetary disks of carbon-enhanced metal-poor (CEMP) stars, possible relics of the early Universe. The chemically anomalous abundance patterns ([C/Fe] $\geq$ 0.7) in this subset of low-mass stars suggest pollution by primordial core-collapsing supernovae (SNe) ejecta that are particularly rich in carbon dust grains.
By comparing the dust-settling timescale in the protoplanetary disks of CEMP stars to the expected disk lifetime (assuming dissipation via photoevaporation), we determine the maximum distance $r_{max}$ from the host CEMP star at which carbon-rich planetesimal formation is possible, as a function of the host star's [C/H] abundance. We then use our linear relation between $r_{max}$ and [C/H], along with the theoretical mass-radius relation derived for a solid, pure carbon planet, to characterize potential planetary transits across host CEMP stars. Given that the related transits are detectable with current and upcoming space-based transit surveys, we suggest initiating an observational program to search for carbon planets around CEMP stars in hopes of shedding light on the question of how early planetary systems may have formed after the Big Bang.
\end{abstract}


\begin{keywords}
cosmology: theory --- early Universe --- planets and satellites: formation ---  planets and satellites: detection --- stars: chemically peculiar --- stars: carbon 
\end{keywords}

\section{Introduction}
The questions of when, where, and how the first planetary systems formed in cosmic history remain crucial to our understanding of structure formation and the emergence of life in the early Universe \citep{2014IJAsB..13..337L}.
In the Cold Dark Matter model of hierarchical structure formation, the first stars are predicted to have formed in dark matter haloes that collapsed at redshifts $z \lesssim$ 50, about 100 million years after the Big Bang \citep{1997ApJ...474....1T,2001PhR...349..125B,2003ApJ...592..645Y,2004ARA&A..42...79B,book}. These short-lived, metal-free, massive first-generation stars ultimately exploded as supernovae (SNe) and enriched the interstellar medium (ISM) with the heavy elements fused in their cores. The enrichment of gas with metals that had otherwise been absent in the early Universe enabled the formation of the first low-mass stars, and perhaps, marked the point at which star systems could begin to form planets \citep{2003Natur.425..812B,2007MNRAS.380L..40F,2008ApJ...672..757C}. In the core accretion model of planet formation (e.g. \citealt{2006RPPh...69..119P,2011ApJ...736...89J}), elements heavier than hydrogen and helium are necessary not only to form the dust grains that are the building blocks of planetary cores, but to extend the lifetime of the protostellar disk long enough to allow the dust grains to grow via merging and accretion to form planetesimals \citep{2005A&A...430.1133K,2009ApJ...704L..75J,2009ApJ...705...54Y,2010MNRAS.402.2735E}.

In the past four decades, a broad search has been launched for low-mass Population II stars in the form of extremely metal-poor sources within the halo of the Galaxy. The HK survey \citep{1985AJ.....90.2089B,1992AJ....103.1987B}, the Hamburg/ESO Survey \citep{1996A&AS..115..227W,2008A&A...484..721C}, the Sloan Digital Sky Survey (SDSS; \citealt{2000AJ....120.1579Y}), and the SEGUE survey \citep{2009AJ....137.4377Y} have all significantly enhanced the sample of metal-poor stars with [Fe/H] $<$ --2.0. Although these iron-poor stars are often referred to in the literature as ``metal-poor" stars, it is critical to note that [Fe/H] does not necessarily reflect a stellar atmosphere's total metal content. The equivalence between `metal-poor" and ``Fe-poor" appears to fall away for stars with [Fe/H] $<$ --3.0 since many of these stars exhibit large overabundances of elements such as C, N, and O; the total mass fractions, $Z$, of the elements heavier than He are therefore not much lower than the solar value in these iron-poor stars. 

Carbon-enhanced metal-poor (CEMP) stars comprise one such chemically anomalous class of stars, with carbon-to-iron ratios [C/Fe] $\geq$ 0.7 (as defined in \citealt{2007ApJ...655..492A,2012ApJ...744..195C,2013ApJ...762...28N}). The fraction of sources that fall into this category increases from $\sim$15-20\% for stars with [Fe/H] $<$ --2.0, to 30\% for [Fe/H] $<$ --3.0, to $\sim$75\% for [Fe/H] $<$ --4.0 \citep{2005ARA&A..43..531B,2013ApJ...762...28N,2015ARA&A..53..631F}. Furthermore, the degree of carbon enhancement in CEMP stars has been shown to notably increase as a function of decreasing metallicity, rising from [C/Fe] $\sim$ 1.0 at [Fe/H] = -1.5 to [C/Fe] $\sim$ 1.7 at [Fe/H] = -2.7. \citep{2012ApJ...744..195C}. Given the significant frequency and level of carbon-excess in this subset of metal-poor Population II stars, the formation of carbon planets around CEMP stars in the early universe presents itself as an intriguing possibility.

From a theoretical standpoint, the potential existence of carbon exoplanets, consisting of carbides and graphite instead of Earth-like silicates, has been suggested by \citet{2005astro.ph..4214K}. Using the various elemental abundances measured in planet-hosting stars, subsequent works have sought to predict the corresponding variety of terrestrial exoplanet compositions expected to exist \citep{2010ApJ...715.1050B,2012ApJ...747L...2C,2012ApJ...760...44C}. Assuming that the stellar abundances are similar to those of the original circumstellar disk, related simulations yield planets with a whole range of compositions, including some that are almost exclusively C and SiC; these occur in disks with C/O $>$ 0.8, favorable conditions for carbon condensation \citep{1975GeCoA..39..389L}. Observationally, there have also been indications of planets with carbon-rich atmospheres, e.g. WASP-12b \citep{2011Natur.469...64M}, and carbon-rich interiors, e.g. 55 Cancri e \citep{2012ApJ...759L..40M}.

In this paper, we explore the possibility of carbon planet formation around the iron-deficient, but carbon-rich subset of low-mass stars, mainly, CEMP stars. In \S2, we discuss the origins of the unique elemental abundance patterns among these C-rich objects and their potential implications for the carbon dust content of the gas from which CEMP stars and their protostellar disks form. Comparing the expected disk lifetime to the dust-settling timescale in these protostellar disks, we then determine the maximum distance from a host CEMP star at which the formation of a carbon-rich planet is possible (\S3). In \S4, we calculate the theoretical mass-radius relation for such a pure carbon planet and present the corresponding depth and duration of its transit across the face of its host CEMP star in \S5. We conclude with a discussion of our findings in \S6. Standard definitions of elemental abundances and ratios are adopted in this paper. For element X, the logarithmic absolute abundance is defined as the number of atoms of element X per 10$^{12}$ hydrogen atoms, $\log{\epsilon(\textrm{X})}=\log_{10}{(N_{\textrm{X}}/N_{\textrm{Y}})}+12.0.$  For elements X and Y, the logarithmic abundance ratio relative to the solar ratio is defined as [X/Y] = $\log_{10}{(N_{\textrm{X}}/N_{\textrm{Y}})}-\log_{10}{(N_{\textrm{X}}/N_{\textrm{Y}})_\odot}$. The solar abundance set is that of \citet{2009ARA&A..47..481A}, with a solar metallicity Z$_\odot$ = 0.0134.

\section{Star-forming environment of CEMP stars}

A great deal of effort has been directed in the literature towards understanding theoretically, the origin of the most metal-poor stars, and in particular, the large fraction that is C-rich. These efforts have been further perturbed by the fact that CEMP stars do not form a homogenous group, but can rather be further subdivided into two main populations \citep{2005ARA&A..43..531B}: carbon-rich stars that show an excess of heavy neutron-capture elements (CEMP-s, CEMP-r, and CEMP-r/s), and carbon-rich stars with a normal pattern of the heavy elements (CEMP-no). In the following sections, we focus on stars with [Fe/H] $\leq$ --3.0,  which have been shown to fall almost exclusively in the CEMP-no subset \citep{2010IAUS..265..111A}. 

A number of theoretical scenarios have been proposed to explain the observed elemental abundances of these stars, though there is no universally accepted hypothesis. 
The most extensively studied mechanism to explain the origin of CEMP-no stars is the mixing and fallback model, where a ``faint" Population III SN explodes, but due to a relatively low explosion energy, only ejects its outer layers, rich in lighter elements (up to magnesium); its innermost layers, rich in iron and heavier elements, fall back onto the remnant and are not recycled in the ISM \citep{2003Natur.422..871U,2005ApJ...619..427U}. 
This potential link between primeval SNe and CEMP-no stars is supported by recent studies which demonstrate that the observed ratio of carbon-enriched to carbon-normal stars with [Fe/H] $<$ --3.0 is accurately reproduced if SNe were the main source of metal-enrichment in the early Universe \citep{2014MNRAS.445.3039D,2014ApJ...791..116C}. 
Furthermore, the observed abundance patterns of CEMP-no stars have been found to be generally well matched by the nucleosynthetic yields of primordial faint SNe \citep{2005ApJ...619..427U,2005Sci...309..451I,2007ApJ...660..516T,2009ApJ...693.1780J,2013ApJ...762...27Y,2014Natur.506..463K,2014ApJ...792L..32I,2014ApJ...794..100M,2015MNRAS.454.4250M,2014ApJ...785...98T,2015A&A...579A..28B}. These findings suggest that most of the CEMP-no stars were probably born out of gas enriched by massive, first-generation stars that ended their lives as Type II SNe with low levels of mixing and a high degree of fallback.

Under such circumstances, the gas clouds which collapse and fragment to form these CEMP-no stars and their protostellar disks may contain significant amounts of carbon dust grains. 
Observationally, dust formation in SNe ejecta has been inferred from isotopic anomalies in meteorites where graphite, SiC, and Si$_3$N$_4$ dust grains have been identified as SNe condensates \citep{1998M&PS...33..549Z}. Furthermore, in situ dust formation has been unambiguously detected in the expanding ejecta of SNe such as SN 1987A \citep{1989LNP...350..164L, 2014ApJ...782L...2I} and SN 1999em \citep{2003MNRAS.338..939E}. The existence of cold dust has also been verified in the supernova remnant of Cassiopeia A by SCUBA's recent submillimeter observations, and a few solar masses worth of dust is estimated to have condensed in the ejecta \citep{2003Natur.424..285D}.

Theoretical calculations of dust formation in primordial core-collapsing SNe have demonstrated the condensation of a variety of grain species, starting with carbon, in the ejecta, where the mass fraction tied up in dust grains grows with increasing progenitor mass \citep{1989ApJ...344..325K,2001MNRAS.325..726T,2003ApJ...598..785N}. \citet{2014ApJ...794..100M,2015MNRAS.454.4250M} consider, in particular, dust formation in weak Population III SNe ejecta, the type believed to have polluted the birth clouds of CEMP-no stars. 
Tailoring the SN explosion models to reproduce the observed elemental abundances of CEMP-no stars, they find that:
(i) for all the progenitor models investigated, amorphous carbon (AC) is the only grain species that forms in significant amounts; this is a consequence of extensive fallback, which results in a distinct, carbon-dominated ejecta composition with negligible amounts of other metals, such as Mg, Si, and Al, that can enable the condensation of alternative grain types;
(ii) the mass of carbon locked into AC grains increases when the ejecta composition is characterized by an initial mass of C greater than the O mass; this is particularly true in zero metallicity supernova progenitors, which undergo less mixing than their solar metallicity counterparts \citep{2009ApJ...693.1780J}; in their stratified ejecta, C-grains are found only to form in layers where C/O $>$ 1; in layers where C/O $<$ 1, all the carbon is promptly locked in CO molecules;
(iii) depending on the model, the mass fraction of dust (formed in SNe ejecta) that survives the passage of a SN reverse shock ranges between 1 to 85\%; this fraction is referred to as the carbon condensation efficiency;
(iv) further grain growth in the collapsing birth clouds of CEMP-no stars, due to the accretion of carbon atoms in the gas phase onto the remaining grains, occurs only if C/O $>$ 1 and is otherwise hindered by the formation of CO molecules.

Besides the accumulation of carbon-rich grains imported from the SNe ejecta, Fischer-Trope-type reactions (FTTs) may also contribute to solid carbon enrichment in the protostellar disks of CEMP-no stars by enabling the conversion of nebular CO and H$_2$ to other forms of carbon \citep{1998M&PS...33..243L,2001M&PS...36...75K}. Furthermore, in carbon-rich gas, the equilibrium condensation sequence changes signifcantly from the sequence followed in solar composition gas where metal oxides condense first. In nebular gas with C/O $\gtrsim$ 1, carbon-rich compounds such as graphite, carbides, nitrides, and sulfides are the highest temperature condensates ($T \approx$ 1200-1600 K) \citep{1975GeCoA..39..389L}. Thus, if planet formation is to proceed in this C-rich gas, the protoplanetary disks of these CEMP-no stars may spawn many carbon planets.

\section{Orbital radii of Potential Carbon Planets }

Given the significant abundance of carbon grains, both imported from SNe ejecta and produced by equilibrium and non-equilibrium mechanisms operating in the C-rich protoplanetary disks, the emerging question is: would these dust grains have enough time to potentially coagulate and form planets around their host CEMP-no stars?

In the core accretion model, terrestrial planet formation is a multi-step process, starting with the aggregation and settling of dust grains in the protoplanetary disk \citep{1993ARA&A..31..129L,2000prpl.conf..533B,2006RPPh...69..119P,2007prpl.conf..639N,2008exop.book...89R,Armitagebook,2011ApJ...736...89J}. In this early stage, high densities in the disk allow particles to grow from submicron-size to meter-size through a variety of collisional processes including Brownian motion, settling, turbulence, and radial migration. The continual growth of such aggregates by coagulation and sticking eventually leads to the formation of kilometer-sized planetesimals, which then begin to interact gravitationally and grow by pairwise collisions, and later by runaway growth \citep{1988mess.book..348W,1993book,1993ARA&A..31..129L}. 
In order for terrestrial planets to ultimately form, these processes must all occur within the lifetime of the disk itself, a limit which is set by the relevant timescale of the physical phenomena that drive disk dissipation. 

A recent study by \citet{2009ApJ...705...54Y} of clusters in the Extreme Outer Galaxy (EOG) provides observational evidence that low-metallicity disks have shorter lifetimes ($<$ 1 Myr) compared to solar metallicity disks ($\sim$ 5-6 Myr). This finding is consistent with models in which photoevaporation by energetic (ultraviolet or X-ray) radiation of the central star is the dominant disk dispersal mechanism. While the opacity source for EUV (extreme-ultraviolet) photons is predominantly hydrogen and is thus metallicity-independent, X-ray photons are primarily absorbed by heavier elements, mainly carbon and oxygen, in the inner gas and dust shells. Therefore, in low metallicity environments where these heavy elements are not abundant and the opacity is reduced, high density gas at larger columns can be ionized and will experience a photoevaporative flow if heated to high enough temperatures \citep{2009ApJ...690.1539G,2010MNRAS.402.2735E}. 

Assuming that photoevaporation is the dominant mechanism through which circumstellar disks lose mass and eventually dissipate, we adopt the metallicity-dependent disk lifetime, derived in \citet{2010MNRAS.402.2735E} using X-ray+EUV models \citep{2009ApJ...699.1639E},
\begin{equation}
t_{disk}\propto Z^{0.77(4-2p)/(5-2p)}
\end{equation}
where $Z$ is the total metallicity of the disk and $p$ is the power-law index of the disk surface density profile ($\Sigma \propto r^{-p}$). A mean power-law exponent of $p \sim$ 0.9 is derived by modeling the spatially resolved emission morphology of young stars at (sub)millimeter wavelengths \citep{2009ApJ...700.1502A,2010ApJ...723.1241A} and the timescale is normalized such that the mean lifetime for disks of solar metallicity is 2 Myr \citep{2010MNRAS.402.2735E}. 
We adopt the carbon abundance relative to solar [C/H] as a proxy for the overall metallicity $Z$ since the opacity, which largely determines the photoevaporation rate, and thus the disk lifetime, is dominated by carbon dust grains in the CEMP-no stars we consider in this paper.

The timescale for planet formation is believed to be effectively set by the time it takes dust grains to settle into the disk midplane. The subsequent process of runaway planetesimal formation, possibly occurring via a series of pairwise collisions, must be quick, since otherwise, the majority of the solid disk material would radially drift towards the host star and evaporate in the hot inner regions of the circumstellar disk \citep{Armitagebook}. 
We adopt the one-particle model of \citet{2005A&A...434..971D} to follow the mass growth of dust grains via collisions as they fall through and sweep up the small grains suspended in the disk.
Balancing the gravitational force felt by a small dust particle at height $z$ above the mid-plane of a disk with the aerodynamic drag (in the Epstein regime) gives a dust settling velocity of 
\begin{equation}
v_{sett}=\frac{dz}{dt}=\frac{3\Omega_K^2zm}{4\rho c_s\sigma_d}
\end{equation}
where $\sigma_d=\pi a^2$ is the cross-section of the dust grain with radius $a$ and $c_s=\sqrt{k_BT(r)/\mu m_H}$ is the isothermal sound speed with $m_H$ being the mass of a hydrogen atom and $\mu$=1.36 being the mean molecular weight of the gas (including the contribution of helium). $\Omega_K=\sqrt{GM_*/r^3}$ is the Keplerian velocity of the disk at a distance $r$ from the central star of mass $M_*$, which we take to be $M_*$ = 0.8 M$_\odot$ as representative of the low-masses associated with CEMP-no stars \citep{2002Natur.419..904C,2015ARA&A..53..631F}. 
The disk is assumed to be in hydrostatic equilibrium with a density given by
\begin{equation}
\rho(z,r)=\frac{\Sigma(r)}{h\sqrt{2\pi}}\exp{\left(-\frac{z^2}{2h^2}\right)}
\end{equation}
where the disk scale height is $h=c_s/\Omega_k$. For the disk surface density $\Sigma(r)$ and temperature $T(r)$ profiles, we adopt the radial power-law distributions fitted to (sub-)millimeter observations of circumstellar disks around young stellar objects \citep{2005ApJ...631.1134A,2009ApJ...700.1502A,2010ApJ...723.1241A},
\begin{equation}
T(r)=200 \textrm{ K} \left(\frac{r}{1 \textrm{ AU}}\right)^{-0.6} 
\end{equation}
\begin{equation}
\Sigma(r)=10^3 \textrm{ g/cm$^2$}\left(\frac{r}{1 \textrm{ AU}}\right)^{-0.9} \,\,.
\end{equation}
Although these relations were observationally inferred from disks with solar-like abundances, we choose to rely on them for our purposes given the lack of corresponding measurements for disks around stars with different abundance patterns.

The rate of grain growth, $dm/dt$, is determined by the rate at which grains, subject to small-scale Brownian motion, collide and stick together as they drift towards the disk mid-plane through a sea of smaller solid particles. If coagulation results from every collision, then the mass growth rate of a particle is effectively the amount of solid material in the volume swept out the particle's geometric cross-section,
\begin{equation}
\frac{dm}{dt}=f_{dg}\rho\sigma_d\left(v_{rel}+\frac{dz}{dt}\right)
\end{equation}
where $dz/dt$ is the dust settling velocity given by equation (2) and
\begin{equation}
v_{rel}=\sqrt{\frac{8k_BT(m_1+m_2)}{\pi m_1m_2}}\approx\sqrt{\frac{8k_BT}{\pi m}}
\end{equation}
is the relative velocity in the Brownian motion regime between grains with masses $m_1=m_2=m$. To calculate the dust-to-gas mass ratio in the disk $f_{dg}$, we follow the approach in \citet{2014ApJ...782...95J} and relate two expressions for the mass fraction of C: (i) the fraction of carbon in the dust, $f_{dg}M_{C,dust}/M_{dust}$, where $M_{dust}$ is the total dust mass and $M_{C,dust}$ is the carbon dust mass ; and (ii) the fraction of carbon in the gas, $\mu_C n_C/\mu n_H$, where $\mu_C$ is the molecular weight of carbon ($\sim12m_p$) and $n_C$ and $n_H$ are the carbon and hydrogen number densities, respectively.

\begin{table}
\centering\begin{minipage}{90mm}
  \caption{Basic data$^a$ for CEMP stars considered in this paper}
  \begin{tabular}{@{}lccccccccc@{}}
  \hline
  \hline
 Star & $\log{g}^b$ & [Fe/H] & [C/Fe] & C/O$^c$ & Source$^d$\\
 \hline
  HE\,0107-5240 & 2.2 & -5.44 & 3.82  & 14.1 & 1,2\\
 SDSS J0212+0137 & 4.0 & -3.57 & 2.26 & 2.6 & 3\\
 SDSS J1742+2531 & 4.0 & -4.77 & 3.60 & 2.2 & 3\\
 G\,77-61 & 5.1 & -4.03 & 3.35 & 12.0 & 4, 5\\
 HE\,2356-0410$^e$ & 2.65 & -3.19 & 2.61 & $>$14.1 & 6\\
\hline
\hline
\end{tabular}
\\
$^a$ Abundances based on one-dimensional LTE model-atmosphere analyses\\
$^b$ Logarithm of the gravitational acceleration at the surface of stars expressed in cm\,s$^{-2}$\\
$^c$ C/O = $N_{\textrm{C}}/N_{\textrm{O}} = 10^{\textrm{[C/O]}+\log{\epsilon(\textrm{C})_\odot}-\log{\epsilon(\textrm{O})_\odot}}$\\
$^d$ \textbf{References:} (1) \citealt{2004ApJ...603..708C}; (2) \citealt{2006ApJ...644L.121C}; (3) \citealt{2015A&A...579A..28B}; (4) \citealt{2005A&A...434.1117P}; (5) \citealt{2007AJ....133.1193B}; (6) \citealt{2014AJ....147..136R}.\\
$^e$ CS\,22957-027
\end{minipage}
\end{table}


\noindent We then assume that a fraction $f_{cond}$ (referred to from now on as the carbon condensation efficiency) of all the carbon present in the gas cloud is locked up in dust, such that
\begin{equation}
f_{cond}\frac{\mu_C n_C}{\mu n_H}=f_{dg}\frac{M_{C,dust}}{M_{dust}} \,\,.
\end{equation}
Since faint Population III SNe are believed to have polluted the birth clouds of CEMP-no stars, and the only grain species that forms in non-negligible amounts in these ejecta is amorphous carbon \citep{2014ApJ...794..100M,2015MNRAS.454.4250M}, we set $M_{dust}=M_{C,dust}$. 
Rewriting equation (8) in terms of abundances relative to the Sun, we obtain
\begin{equation}
f_{dg}=f_{cond}\frac{\mu_C}{\mu}10^{\textrm{[C/H]}+\log{\epsilon(C)_\odot}-12}
\end{equation}
where $\log{\epsilon(C)_\odot}$=8.43$\pm$0.05 \citep{2009ARA&A..47..481A} is the solar carbon abundance.

For a specified metallicity [C/H] and radial distance $r$ from the central star, we can then estimate the time it takes for dust grains to settle in the disk by integrating equations (2) and (6) from an initial height of $z(t=0)=4h$ with an initial dust grain mass  of $m(t=0)=4\pi a_{init}^3\rho_d/3$. The specific weight of dust is set to $\rho_d$=2.28 g cm$^{-3}$, reflecting the material density of carbon grains expected to dominate the circumstellar disks of CEMP-no stars. The initial grain size $a_{init}$ is varied between 0.01 and 1 $\mu$m to reflect the range of characteristic radii of carbon grains found when modeling CEMP-no star abundance patterns \citep{2014ApJ...794..100M}. Comparing the resulting dust-settling timescale to the disk lifetime given by equation (1) for the specified metallicity, we can then determine whether there is enough time for carbon dust grains to settle in the mid-plane of the disk and there undergo runaway planetesimal formation before the disk is dissipated by photoevaporation. For the purposes of this simple model, we neglected possible turbulence in the disk which may counteract the effects of vertical settling, propelling particles to higher altitudes and thus preventing them from fully settling into the disk mid-plane \citep{Armitagebook}. We have also not accounted for the effects of radial drift, which may result in the evaporation of solid material in the hot inner regions of the circumstellar disk.


\begin{figure}
\hspace{-1.3cm}\includegraphics[width=105mm,height=80mm]{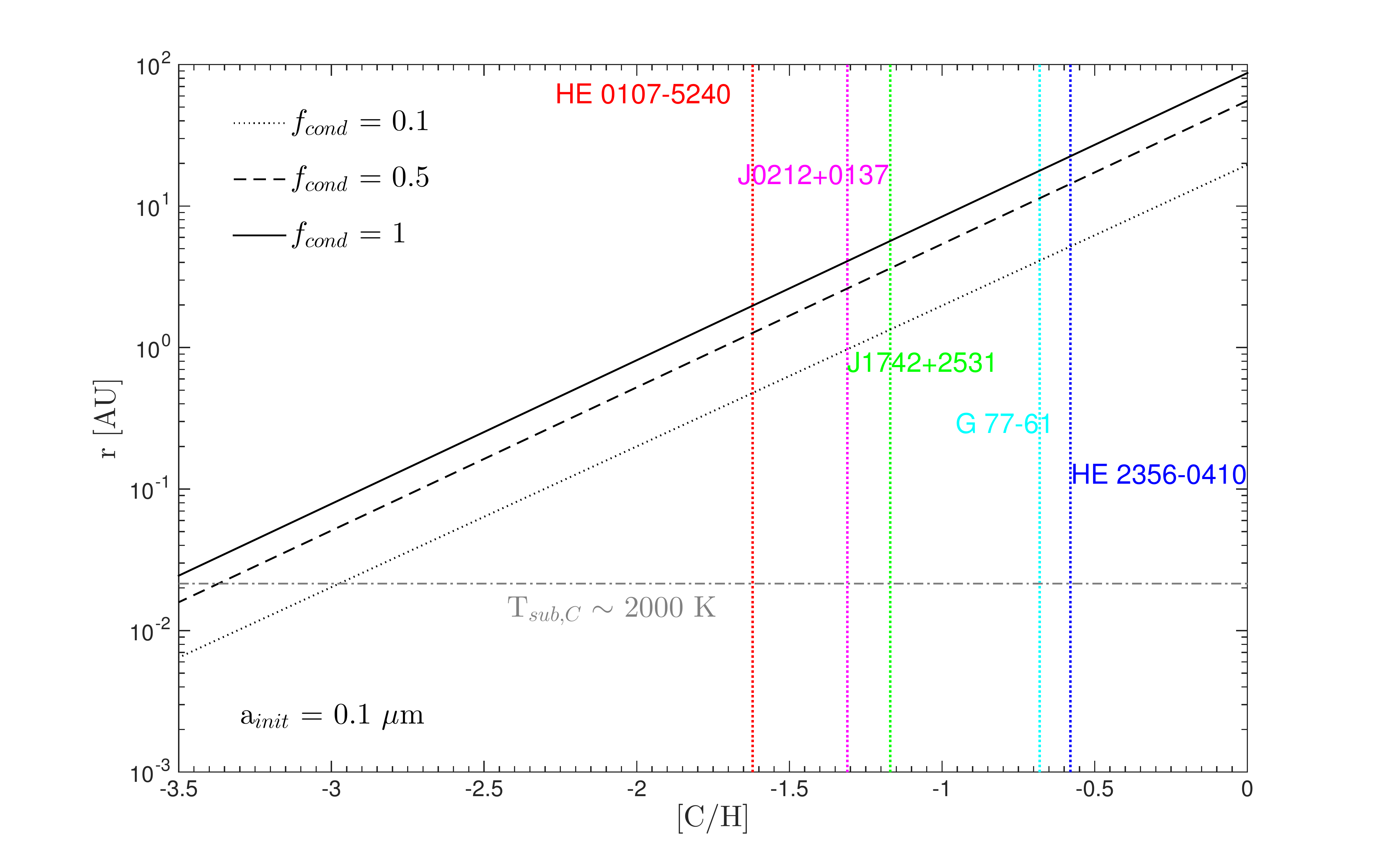}
\caption{The maximum distance $r_{max}$ from the host star out to which planetesimal formation is possible as a function of the star's metallicity, expressed as the carbon abundance relative to that of the Sun, [C/H]. The dotted, dashed, and solid black curves correspond to the results obtained assuming carbon condensation efficiencies of 10\%, 50\%, and 100\%, respectively, and an initial grain size of $a_{init}$ = 0.1 $\mu$m. The gray dash-dotted curve corresponds to the distance at which the disk temperature approaches the sublimation temperature of carbon dust grains, $T_{sub,\textrm{C}} \sim$ 2000 K; the formation of carbon planetesimals will therefore be suppressed at distances that fall below this line, $r \lesssim$ 0.02 AU. The colored vertical lines represent various observed CEMP stars with measured carbon abundances, [C/H].}
\end{figure}


As the dust settling timescale is dependent on the disk surface density $\Sigma(r)$ and temperature $T(r)$, we find that for a given metallicity, [C/H], there is a maximum distance $r_{max}$ from the central star out to which planetesimal formation is possible. At larger distances from the host star, the dust settling timescale exceeds the disk lifetime and so carbon planets with semi-major axes $r > r_{max}$ are not expected to form. 
A plot of the maximum semi-major axis expected for planet formation around a CEMP-no star as a function of the carbon abundance relative to the Sun [C/H] is shown in Figure 1 for carbon condensation efficiencies ranging between $f_{cond}$ = 0.1 and 1. As discovered in \citet{2012ApJ...751...81J} where the critical iron abundance for terrestrial planet formation is considered as a function of the distance from the host star, we find a linear relation between [C/H] and $r_{max}$,
\begin{equation}
\textrm{[C/H]}=\log{\left(\frac{r_{max}}{1 \textrm{ AU}}\right)}-\alpha
\end{equation}
where $\alpha$ = 1.3, 1.7, and 1.9 for $f_{cond}$ = 0.1, 0.5, and 1, respectively, assuming an initial grain size of $a_{init}$ = 0.1 $\mu$m. These values for $\alpha$ change by less than 1\% for smaller initial grain sizes, $a_{init}$ = 0.01 $\mu$m, and by no more than 5\% for larger initial grain sizes $a_{init}$ = 1 $\mu$m; given this weak dependence on $a_{init}$, we only show our results for a single initial grain size of $a_{init}$ = 0.1 $\mu$m. The distance from the host star at which the temperature of the disk approaches the sublimation temperature of carbon dust, $T_{sub,C} \sim$ 2000 K \citep{2011EP&S...63.1067K},  is depicted as well (dash-dotted gray curve). At distances closer to the central star than $r \simeq$ 0.02 AU, temperatures well exceed the sublimation temperature of carbon grains; grain growth and subsequent carbon planetesimal formation are therefore quenched in this inner region.

Figure 1 shows lines representing various observed CEMP stars with measured carbon abundances, mainly, HE\,0107-5240 \citep{2002Natur.419..904C,2004ApJ...603..708C}, SDSS J0212+0137 \citep{2015A&A...579A..28B}, SDSS J1742+2531 \citep{2015A&A...579A..28B}, G\,77-61 \citep{1977ApJ...216..757D,2005A&A...434.1117P,2007AJ....133.1193B}, and HE\,2356-0410 \citep{1997ApJ...489L.169N,2014AJ....147..136R}. These stars all have iron abundances (relative to solar) [Fe/H] $<$ -3.0, carbon abundances (relative to solar) [C/Fe] $>$ 2.0, and carbon-to-oxygen ratios C/O $>$ 1. This latter criteria maximizes the abundance of solid carbon available for planet formation in the circumstellar disks by optimizing carbon grain growth both in stratified SNe ejecta and later, in the collapsing molecular birth clouds of these stars. It also advances the possibility of carbon planet formation by ensuring that planet formation proceeds by a carbon-rich condensation sequence in the protoplanetary disk. 
SDSS J0212+0137 and HE\,2356-0410 have both been classified as CEMP-no stars, with measured barium abundances [Ba/Fe] $<$ 0 (as defined in \citealt{2005ARA&A..43..531B}); the other three stars are Ba-indeterminate, with only high upper limits on [Ba/Fe], but are believed to belong to the CEMP-no subclass given their light-element abundance patterns. 
The carbon abundance, [C/H], dominates the total metal content of the stellar atmosphere in these five CEMP objects, contributing more than 60\% of the total metallicity in these stars. 
A summary of the relevant properties of the CEMP stars considered in this analysis can be found in Table 1. 
We find that carbon planets may be orbiting iron-deficient stars with carbon abundances [C/H] $\sim$ -0.6, such as HE\,2356-0410, as far out as $\sim$ 20 AU from their host star in the case where $f_{cond}$ = 1. Planets forming around stars with less carbon enhancement, i.e. HE\,0107-5240 with [C/H] $\sim$ -1.6, are expected to have more compact orbits, with semi-major axes $r <$ 2 AU. If the carbon condensation efficiency is only 10\%, the expected orbits grow even more compact, with maximum semi-major axes of $\sim$ 5 and 0.5 AU, respectively.

\section{Mass-Radius Relationship for Carbon Planets}
Next we present the relationship between the mass and radius of carbon planets that we have shown may theoretically form around CEMP-no stars. These mass-radius relations have already been derived in the literature for a wide range of rocky and icy exoplanet compositions \citep{1969ApJ...158..809Z,2004Icar..169..499L,2006Icar..181..545V,2007ApJ...659.1661F,2007ApJ...669.1279S}. Here, we follow the approach of \citet{1969ApJ...158..809Z} and solve the three canonical equations of internal structure for solid planets,
\begin{enumerate}
\item mass conservation
\begin{equation}
\frac{dm(r)}{dr}=4\pi r^2 \rho(r) \,,
\end{equation}
\item hydrostatic equilbrium
\begin{equation}
\frac{dP(r)}{dr}=-\frac{Gm(r)\rho(r)}{r^2} \,,\,\,\textrm{and}
\end{equation}
\item the equation of state (EOS)
\begin{equation}
P(r) = f\left(\rho(r),T(r)\right) \,,
\end{equation}
\end{enumerate}
where $m(r)$ is the mass contained within radius $r$, $P(r)$ is the pressure, $\rho(r)$ is the density of the spherical planet, and $f$ is the unique equation of state (EOS) of the material of interest, in this case, carbon.

Carbon grains in circumstellar disks most likely experience many shock events during planetesimal formation which may result in the modification of their structure. The coagulation of dust into clumps, the fragmentation of the disk into clusters of dust clumps, the merging of these clusters into $\sim$ 1 km planetesimals, the collision of planetesimals during the accretion of meteorite parent bodies, and the subsequent collision of the parent bodies after their formation all induce strong shock waves that are expected to chemically and physically alter the materials \citep{Mimurabook}.
Subject to these high temperatures and pressures, the amorphous carbon grains polluting the protoplanetary disks around CEMP stars are expected to undergo graphitization and may even crystallize into diamond \citep{1987ApJ...319L.109T,1988JMatS..23..422O,1996A&A...315..222P,Takai2003}.
In our calculations, the equation of state at low pressures, $P \leq$ 14 GPa, is set to the third-order finite strain Birch-Murnagham EOS (BME; \citealt{1947PhRv...71..809B,Poirierbook}) for graphite,
\begin{equation}
P=\frac{3}{2}K_0\left(\eta^{7/3}-\eta^{5/3}\right)\left[1+\frac{3}{4}\left(K_0^'-4\right)\left(\eta^{2/3}-1\right)\right]
\end{equation}
where $\eta=\rho/\rho_0$ is the compression ratio with respect to the ambient density, $\rho_0$, $K_0$ is the bulk modulus of the material, and $K_0^'$ is the pressure derivative. Empirical fits to experimental data yields a BME EOS of graphite ($\rho_0$ = 2.25 g cm$^{-3}$) with parameters $K_0$ = 33.8 GPa and  $K_0^'$ = 8.9 \citep{1989PhRvB..3912598H}. At 14 GPa, we incorporate the phase transition from graphite to diamond \citep{1976Natur.259...38N,1989PhRvB..3912598H} and adopt the Vinet EOS \citep{1987JGR....92.9319V,1989JPCM....1.1941V},
\begin{equation}
P=3K_0\eta^{2/3}\left(1-\eta^{-1/3}\right)\exp{\left[\frac{3}{2}\left(K_0^'-1\right)\left(1-\eta^{-1/3}\right)\right]}
\end{equation}
with $K_0$ = 444.5 GPa and  $K_0^'$ = 4.18 empirically fit for diamond, $\rho_0$ = 3.51 g cm$^{-3}$ \citep{2008PhRvB..77i4106D}. (As pointed out in \citet{2007ApJ...669.1279S}, the BME EOS is not fit to be extrapolated to high pressures since it is derived by expanding the elastic potential energy as a function of pressure keeping only the lowest order terms.) Finally, at pressures $P \gtrsim$ 1300 GPa where electron degeneracy becomes increasingly important, we use the Thomas-Fermi-Dirac (TFD) theoretical EOS (\citealt{1967PhRv..158..876S}; equations (40)-(49)), which intersects the diamond EOS at $P \sim$ 1300 GPa. 
Given that the full temperature-dependent carbon EOSs are either undetermined or dubious at best, all three EOSs adopted in this work are room-temperature EOSs for the sake of practical simplification.


\begin{figure}
\hspace{-.7cm}\includegraphics[width=105mm,height=80mm]{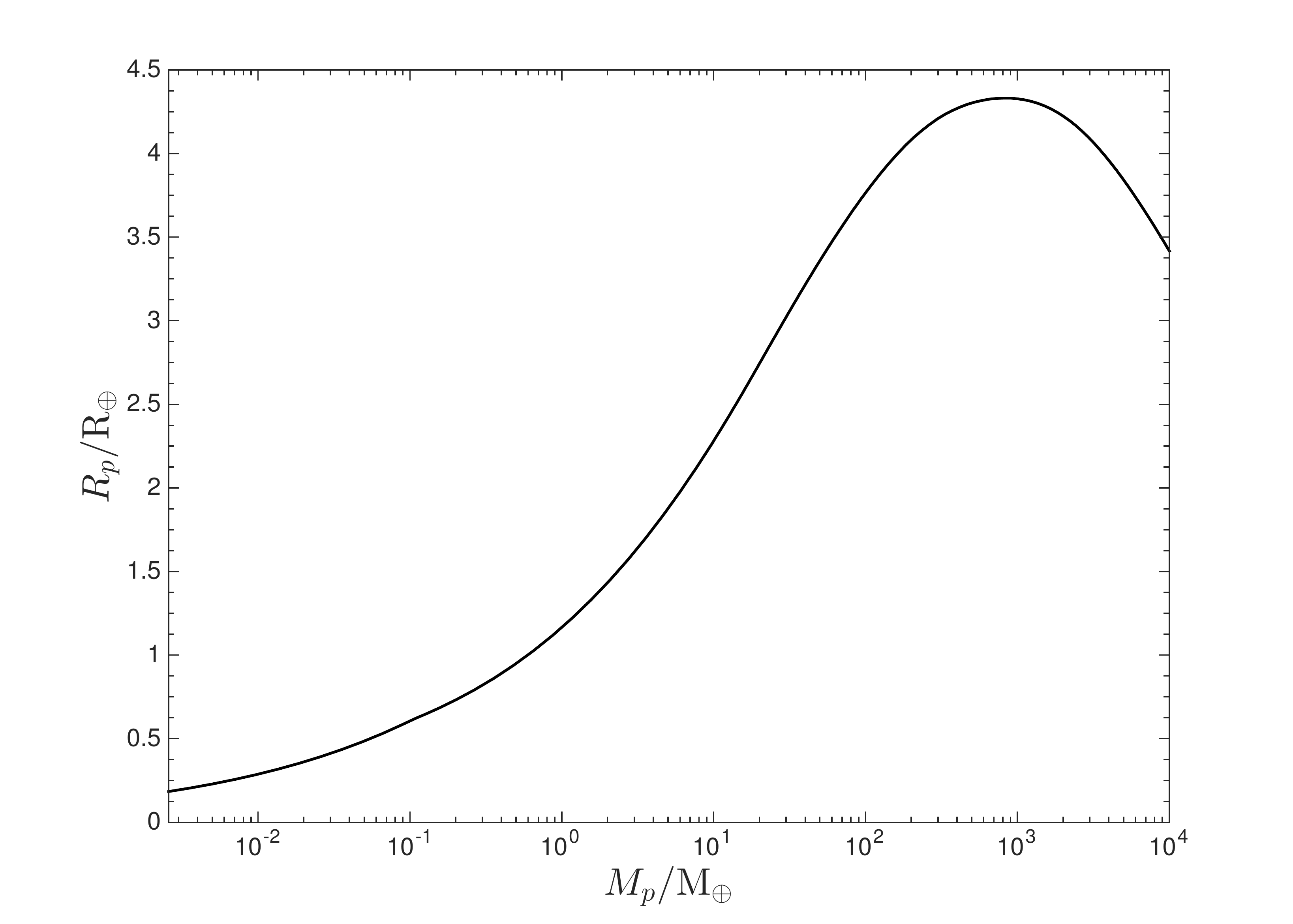}
\caption{Mass-radius relation for solid homogenous, pure carbon planet}
\end{figure}


Using a fourth-order Runge-Kutta scheme, we solve the system of equations simultaneously, numerically integrating equations (11) and (12) begining at the planet's center with the inner boundary conditions $M(r=0)$ = 0 and $P(r=0)$ = $P_{\textrm{central}}$, where $P_{\textrm{central}}$ is the central pressure. The outer boundary condition $P(r=R_p)$ = 0 then defines the planetary radius $R_p$ and total planetary mass $M_p = m(r=R_p)$. Integrating these equations for a range of $P_{\textrm{central}}$, with the appropriate EOS, $P=P(\rho)$, to close the system of equations, yields the mass-radius relationship for a given composition. We show this mass-radius relation for a purely solid carbon planet in Figure 2. We find that for masses $M_p \lesssim$ 800 M$_\oplus$, gravitational forces are small compared with electrostatic Coulomb forces in hydrostatic equilibrium and so the planet's radius increases with increasing mass, $R_p \propto M_p^{1/3}$. However, at larger masses, the electrons are pressure-ionized and the resulting degeneracy pressure becomes significant, causing the planet radius to become constant and even decrease for increasing mass, $R_p \propto M_p^{-1/3}$ \citep{Hubbardbook}. 
Planets which fall within the mass range 500 $\lesssim M_p \lesssim$ 1300 M$_\oplus$, where the competing effects of Coulomb forces and electron degeneracy pressure cancel each other out, are expected to be approximately the same size, with $R_p \simeq$ 4.3 R$_\oplus$, the maximum radius of a solid carbon planet. (In the case of gas giants, the planet radius can increase due to accretion of hydrogen and helium.)

Although the mass-radius relation illustrated in Figure 2 may alone not be enough to confidently distinguish a carbon planet from a water or silicate planet, the unique spectral features in the atmospheres of these carbon planets may provide the needed fingerprints. At high temperatures ($T \gtrsim$ 1000 K), the absorption spectra of massive ($M \sim$ 10 - 60 M$_\oplus$) carbon planets are expected to be dominated by CO, in contrast with the H$_2$O-dominated spectra of hot massive planets with solar-composition atmospheres
\citep{2005astro.ph..4214K}.
The atmospheres of low-mass ($M \lesssim$ 10 M$_\oplus$) carbon planets are also expected to be differentiable from their solar-composition counterparts due to their abundance of CO and CH$_4$, and lack of oxygen-rich gases like CO$_2$, O$_2$, and O$_3$
\citep{2005astro.ph..4214K}.
Furthermore, carbon planets of all masses at low temperatures are expected to accommodate hydrocarbon synthesis in their atmospheres; stable long-chain hydrocarbons are therefore another signature feature that can help observers distinguish the atmospheres of cold carbon planets and more confidently determine the bulk composition of a detected planet
\citep{2005astro.ph..4214K}.

\section{Transit Properties}
The detection of theoretically proposed carbon planets around CEMP stars will provide us with significant clues regarding how early planet formation may have started in the Universe. 
While direct detection of these extrasolar planets remains difficult given the low luminosity of most planets, techniques such as the transit method are often employed to indirectly spot exoplanets and determine physical parameters of the planetary system.
When a planet ``transits" in front of its host star, it partially occludes the star and causes its' observed brightness to drop by a minute amount. If the host star is observed during one of these transits, the resulting dip in its measured light curve can yield information regarding the relevant sizes of the star and the planet, the orbital semi-major axis, and the orbital inclination, among other characterizing properties.


\begin{figure}
\hspace{-1cm}\includegraphics[width=105mm,height=80mm]{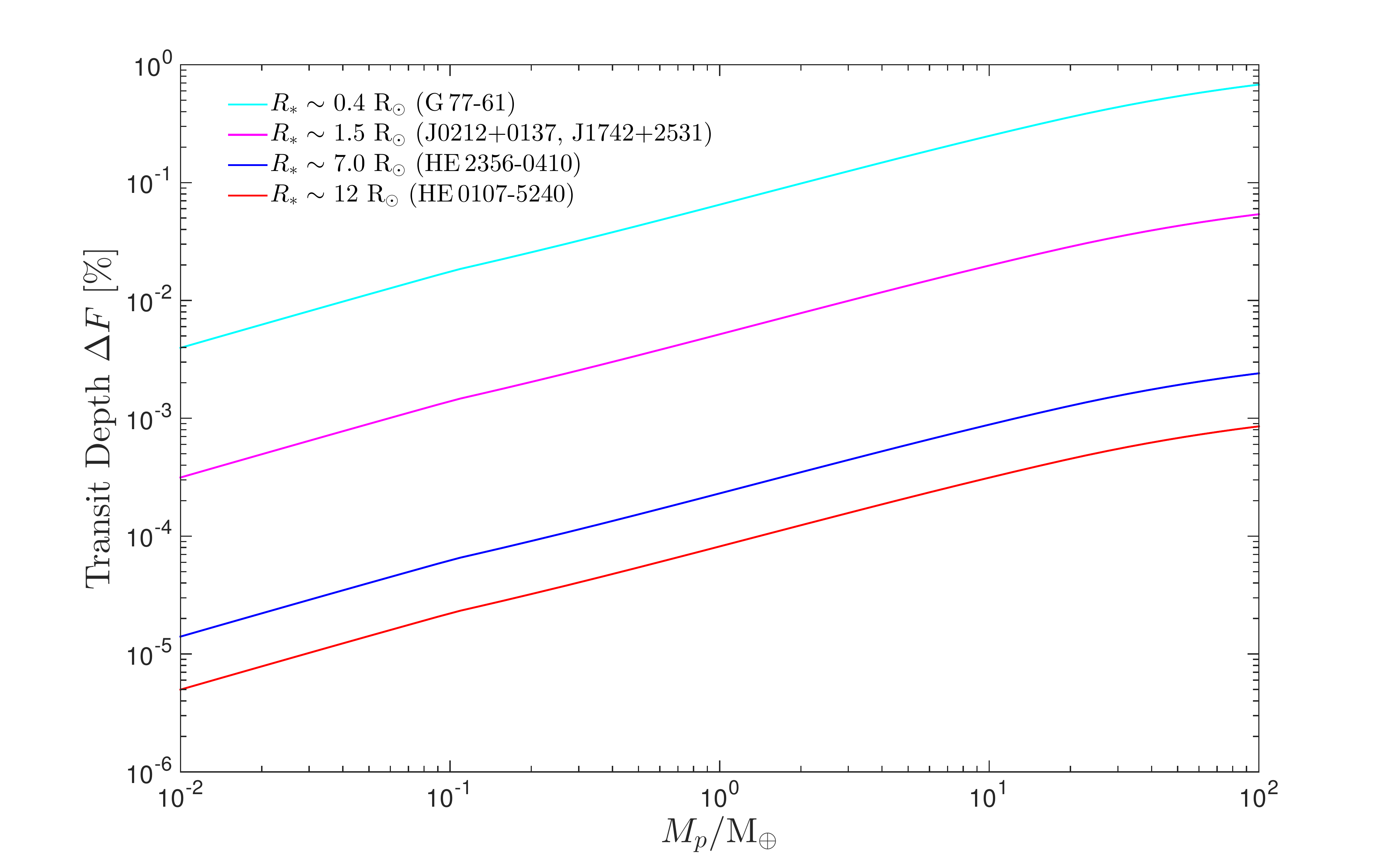}
\caption{Transit depth as a function of the planetary mass (for a solid carbon planet transit) for different host stellar radii with an assumed stellar mass of M$_*$ = 0.8 M$_\odot$. These particular stellar radii were chosen to correspond to the stellar radii of the CEMP stars considered in this paper, mainly HE\,0107-5240 (red), G\,77-61(cyan), HE\,2356-0410 (blue), SDSS\,J0212+0137 and SDSS\,J1742+2531 (magenta), where the last two CEMP objects have the same measured surface gravity $\log{g} = 4.0$.}
\end{figure}


A planetary transit across a star is characterized by three main parameters: the fractional change in the stellar brightness, the orbital period, and the duration of the transit \citep{1996Ap&SS.241..111B}.
The fractional change in brightness is referred to as the transit depth, $\Delta F$ (with a total observed flux $F$), and is simply  defined as the ratio of the planet's area  to the host star's area \citep{2003ApJ...585.1038S},
\begin{equation}
\Delta F = \frac{F_{\textrm{no transit}}-F_{\textrm{transit}}}{F_{\textrm{no transit}}}=\left(\frac{R_p}{R_*}\right)^2 \,.
\end{equation}
Given the stellar radius $R_*$, measurements of the relative flux change $\Delta F$ yield estimates of the size of the planet $R_p$, and the corresponding planetary mass $M_p$ if the mass-radius relation for the planet is known. Using the $R_p$-$M_p$ relation derived in \S4, we illustrate in Figure 3 how the transit depth varies as a function of planetary mass in the case where a pure carbon planet transits across the face of its host CEMP star. Curves are shown for each of the five CEMP stars considered in this study, where we assume a stellar mass of $M_*$ = 0.8 M$_\odot$ (representative of the low masses associated with old, iron-poor stellar objects) and derive the stellar radii using the stellar surface gravities $g$ listed in Table 1, $R_* = \sqrt{GM_*/g}$.


\begin{figure}
\hspace{-0.7cm}\includegraphics[width=105mm,height=80mm]{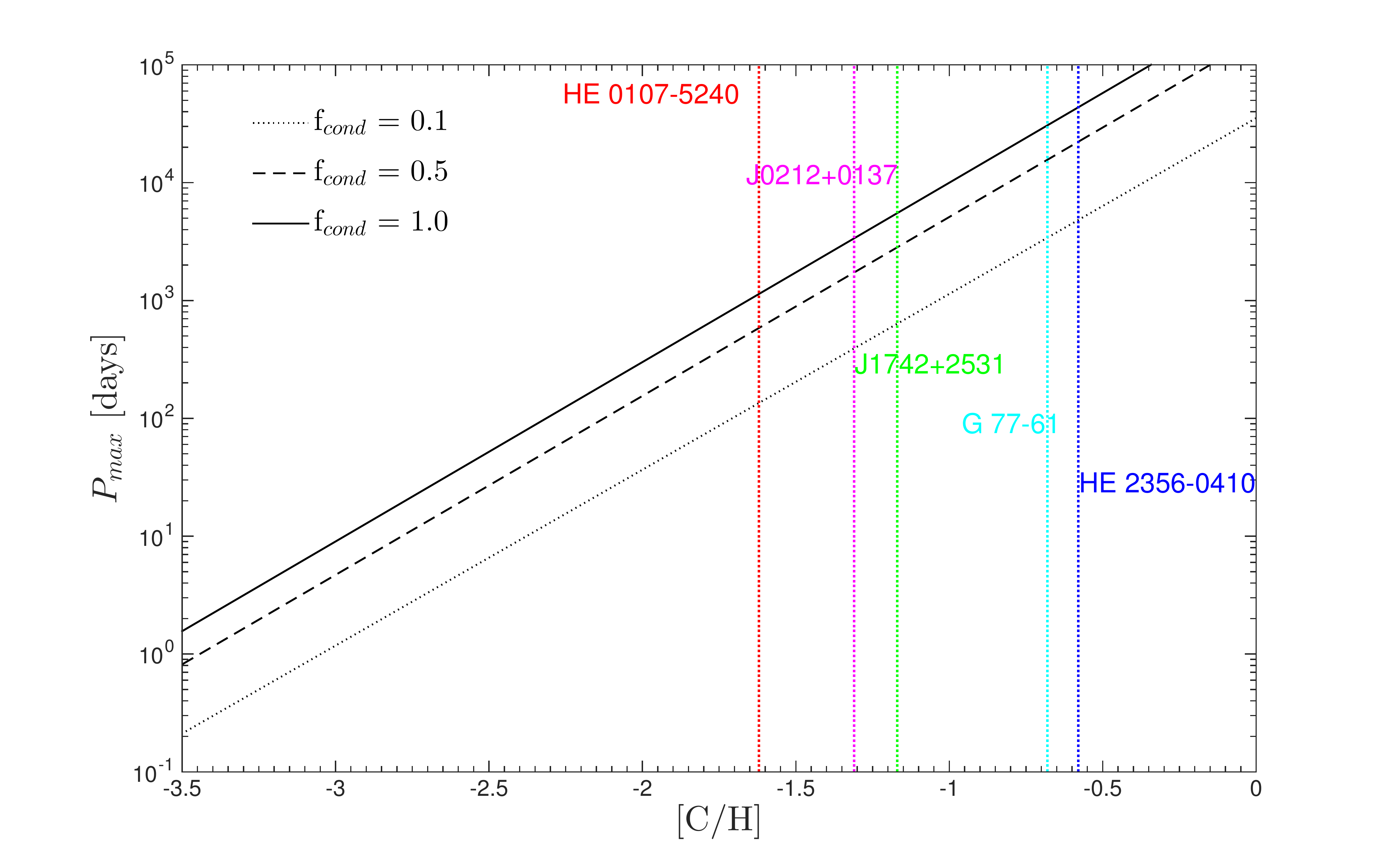}
\caption{Maximum orbital period of a carbon planet transiting across its host CEMP star as a function of the star's metallicity, expressed as the carbon abundance relative to that of the Sun, [C/H].
The dotted, dashed, and solid black curves denote the results obtained assuming carbon condensation efficiencies of 10\%, 50\%, and 100\% in the parent CEMP star with mass $M_*$ = 0.8 M$_\odot$. The colored vertical lines represent the five CEMP stars considered in this paper with measured carbon abundances [C/H].}
\end{figure}


As can be seen in Figure 3, the relative change in flux caused by an Earth-mass carbon planet transiting across its host CEMP star ranges from $\sim$0.0001\% for a host stellar radius of $R_* \sim$ 10 R$_\odot$ to $\sim$0.01\% for a solar-sized stellar object. These shallow transit depths are thus expected to evade detection by ground-based transit surveys, which are generally limited in sensitivity to fractional flux changes on the order of 0.1\% \citep{2009IAUS..253..319B}. 
To push the limits of detection down to smaller, low-mass terrestrial planets requires space-based transit surveys that continuously monitor a large number of potential host stars over several years and measure their respective transit light curves.
There are a number of ongoing, planned, and proposed space missions committed to this cause, including CoRot (COnvection ROtation and planetary Transits), Kepler, PLATO (PLAnetary Transits and Oscillations of stars), TESS (Transiting Exoplanet Survey Satellite), and ASTrO (All Sky Transit Observer), which are expected to achieve precisions as low as 20-30 ppm (parts per million) \citep{2009IAUS..253..319B,2015ARA&A..53..409W}. With the ability to measure transit depths as shallow as $\Delta F \sim$ 0.001\%, these space transit surveys offer a promising avenue towards detecting the planetary systems that may have formed around CEMP stars.


\begin{figure*}
\hspace{-0.8cm}\includegraphics[width=200mm,height=100mm]{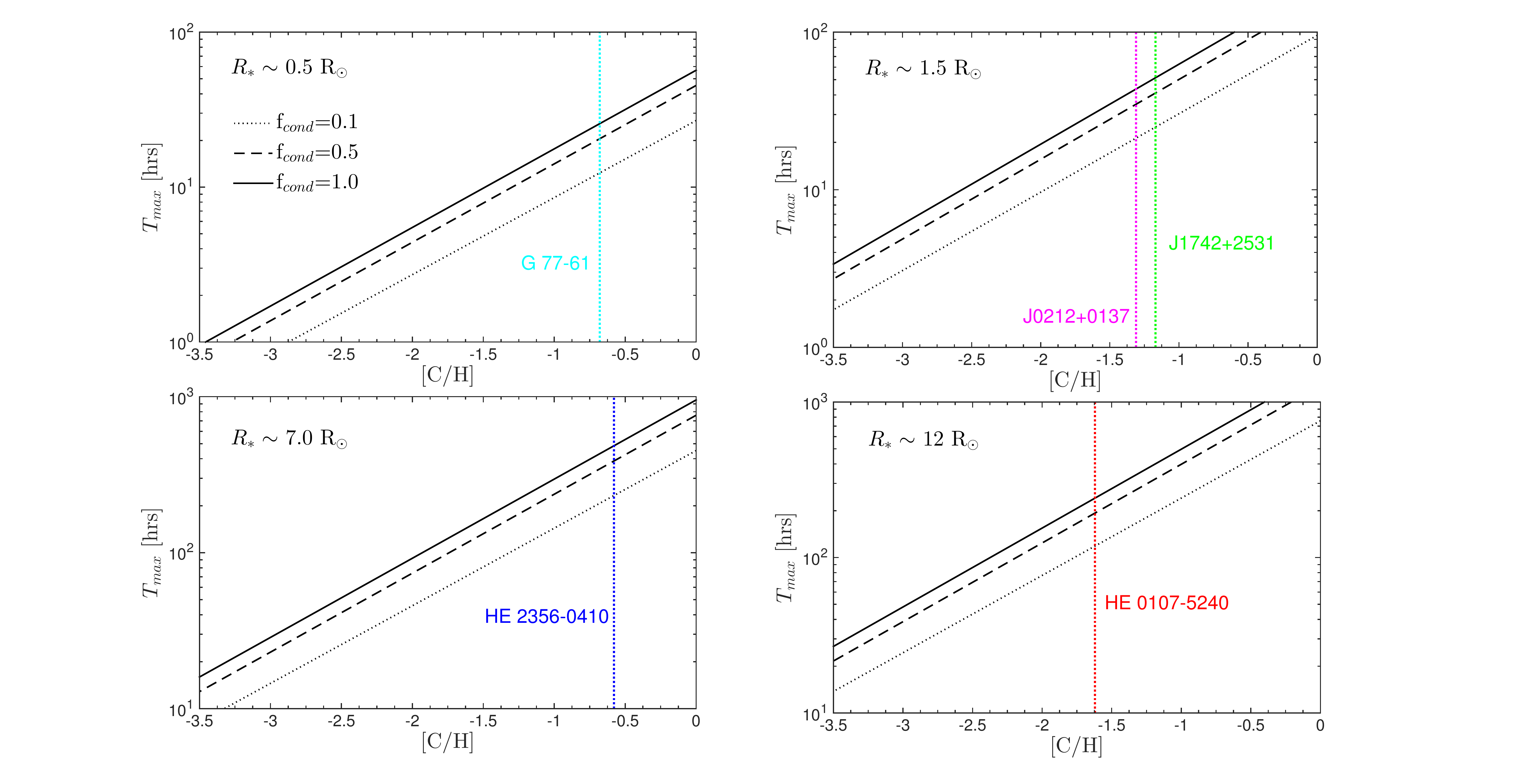}
\caption{Maximum total transit duration of a carbon planet across its host CEMP star as a function of the star's metallicity, expressed as the carbon abundance relative to that of the Sun, [C/H].
The dotted, dashed, and solid black curves denote the results obtained assuming carbon condensation efficiencies of 10\%, 50\%, and 100\%.
 in the parent CEMP star. The colored vertical lines represent the five CEMP stars considered in this paper with measured carbon abundances [C/H] and stellar radii derived using $R_*=\sqrt{GM_*/g}$ with mass $M_*$ = 0.8 M$_\odot$}.
\end{figure*}


The orbital period of a planet $P$, which can be determined if consecutive transits are observed, is given by Kepler's third law in the case of a circular orbit,
\begin{equation}
P^2=\frac{4\pi^2 a^3}{G(M_*+M_p)}\simeq\frac{4\pi^2 a^3}{GM_*}
\end{equation}
where $a$ is the orbital semi-major axis and the planetary mass is assumed to be negligible relative to the stellar mass, $M_p  \ll M_*$, in the second equality. Given the relation we derived in equation (10) between the metallicity [C/H] and the maximum semi-major axis allowed for a planet orbiting a CEMP star, the maximum orbital period of the planet can be expressed as a function of the metallicity of the host CEMP star ($M_*$ = 0.8 M$_\odot$),
\begin{equation}
P_{\textrm{max}}\simeq 365.25 \frac{10^{\frac{3}{2}(\textrm{[C/H]}+\alpha)}}{\sqrt{M_*/\textrm{M}_\odot}} \textrm{ days} 
\end{equation}
where $\alpha \simeq$ 1.3, 1.7, and 1.9 for carbon condensation efficiencies of 10\%, 50\%, and 100\%. As can be seen in Figure 4, CEMP stars with higher carbon abundances [C/H], i.e. G\,77-61 and HE\,2356-0410, and larger efficiencies for carbon dust condensation $f_{cond}$, can host planets with wider orbits and slower rotations, resulting in transits as infrequently as once every couple hundred years. Conversely, carbon planets orbiting relatively less carbon-rich CEMP stars, like SDSS\,J0212+0137, are expected to have higher rates of transit reoccurrence, completing rotations around their parent stars every $\sim$ 1-10 years.
These shorter period planets therefore have a much higher probability of producing an observable transit.

The maximum duration of the transit, $T$, can also be expressed as a function of the metallicity [C/H] of the parent CEMP star. For transits across the center of a star, the total duration is given by,
\begin{equation}
T\simeq2R_*\sqrt{\frac{a}{GM_*}}
\end{equation}
with the assumption that $M_p\ll M_*$ and $R_p\ll R_*$. Once again, using the relation from equation (10) to express the maximum orbital distance from the host star in terms of the star's carbon abundance, we find that the maximum transit duration of a carbon planet across its parent CEMP star ($M_*$ = 0.8 M$_\odot$) is
\begin{equation}
T_{max}\simeq 13\,\frac{R_*}{\textrm{R}_\odot}\sqrt{\frac{10^{\textrm{[C/H]}+\alpha}}{M_*/\textrm{M}_\odot}} \textrm{  hrs}\,\,.
\end{equation}
In Figure 5, these maximum durations are shown as a function of [C/H] for the various stellar radii associated with the CEMP stars we consider in this paper. Transits across CEMP stars with larger radii and higher carbon abundances are expected to take much longer.
While the total transit duration across SDSS\,J0212+0137 and SDSS\,J1742+2531 with $R_*\sim$ 1.5 R$_\odot$ and metallicities of [C/H] $\sim$ -1.3 -- -1.2, is at most $\sim$ 1-2 days, transits across HE\,0107-5240 ($R_*\sim$ 12 R$_\odot$, [C/H]$\sim$ -1.6) can take up to 2 weeks ($f_{cond}$ = 1). 
In general, the geometric probability of a planet passing between the observer and the planet's parent star increases with stellar radius and decreases with orbital radius, $p_t \simeq R_*/a$ \cite{2007MNRAS.380.1488K}. Therefore, focusing on CEMP stars, such as HE\,0107-5240 and HE\,2356-0410, with large stellar radii increases the observer's chance of spotting transits and detecting a planetary system.

\section{Discussion}
We explored in this paper the possibility of carbon planet formation around the iron-deficient, carbon-rich subset of low-mass stars known as CEMP stars. The observed abundance patterns of CEMP-no stars suggest that these stellar objects were probably born out of gas enriched by massive first-generation stars that ended their lives as Type II SNe with low levels of mixing and a high degree of fallback.
The formation of dust grains in the ejecta of these primordial core-collapsing SNe progenitors has been observationally confirmed and theoretically studied. In particular, amorphous carbon is the only grain species found to condense and form in non-negligible amounts in SN explosion models that are tailored to reproduce the abundance patterns measured in CEMP-no stars.
Under such circumstances, the gas clouds which collapse and fragment to form CEMP-no stars and their protoplanetary disks may contain significant amounts of carbon dust grains imported from SNe ejecta.
The enrichment of solid carbon in the protoplanetary disks of CEMP stars may then be further enhanced by Fischer-Trope-type reactions and carbon-rich condensation sequences, where the latter occurs specifically in nebular gas with C/O $\gtrsim$ 1.

For a given metallicity [C/H] of the host CEMP star, the maximum distance out to which planetesimal formation is possible can then be determined by comparing the dust-settling timescale in the protostellar disk to the expected disk lifetime. Assuming that disk dissipation is driven by a metallicity-dependent photoevaporation rate, we find a linear relation between [C/H] and the maximum semi-major axis of a carbon planet orbiting its host CEMP star.
Very carbon-rich CEMP stars, such as G\,77-61 and HE\,2356-0410 with [C/H] $\simeq$ -0.7 -- -0.6, can host carbon planets with semi-major axes as large $\sim$ 20 AU for 100\% carbon condensation efficiencies; this maximum orbital distance reduces to $\sim$ 5 AU when the condensation efficiency drops by an order of magnitude.
In the case of the observed CEMP-no stars HE\,0107-5240, SDSS\,J0212+0137, and SDSS\,J1742+2531, where the carbon abundances are in the range [C/H] $\simeq$ -1.6 -- -1.2, we expect more compact orbits, with maximum orbital distances $r_{max} \simeq$ 2, 4, and 6 AU, respectively, for $f_{cond}$ = 1 and $r_{max} \simeq$ 0.5 - 1 AU for $f_{cond}$ = 0.1.

We then use the linear relation found between [C/H] and $r_{max}$ (\S3), along with the theoretical mass-radius relation derived for a solid, pure carbon planet (\S4), to compute the three observable characteristics of planetary transits: the orbital period, the transit depth, and the transit duration. We find that the relative change in flux, $\Delta F$, caused by an Earth-mass carbon planet transiting across its host CEMP star ranges from $\sim$ 0.0001\% for a stellar radius of $R_* \sim$ 10 R$_\odot$ to $\sim$ 0.01\% for a solar-sized stellar host. 
While the shallow transit depths of Earth-mass carbon planets around HE\,0107-5240 and HE\,2356-0410 may evade detection, current and future space-based transit surveys promise to achieve the precision levels ($\Delta F \sim$ 0.001\%) necessary to detect planetary systems around CEMP stars such as SDSS\,J0212+0137, SDSS\,J1742+2531, and G\,77-61.

Short orbital periods and long transit durations are also key ingredients in boosting the probability of transit detection by observers. G\,77-61 is not an optimal candidate in these respects since given its large carbon abundance ([C/Fe] $\sim$ 3.4), carbon planets may form out to very large distances and take up to a century to complete an orbit around the star for $f_{cond}$ = 1 ($P_{max} \sim$ 10 years for 10\% carbon condensation efficiency). The small stellar radius, $R_* \sim$ 0.5 R$_\odot$, also reduces chances of spotting the transit since the resulting transit duration is only $\sim$ 30 hours at most. Carbon planets around larger CEMP stars with an equally carbon-rich protoplanetary disk, such as HE\,2356-0410 ($R_* \sim$ 7 R$_\odot$), have a better chance of being spotted, with transit durations lasting up to $\sim$ 3 weeks. The CEMP-stars SDSS\,J0212+0137, and SDSS\,J1742+2531 are expected to host carbon planets with much shorter orbits, $P_{max} \sim$ 16 years for 100\% condensation efficiency ($P_{max} \sim$ 1 year for $f_{cond}$ = 0.1), and transit durations that last as long as $\sim$ 60 hours. If the ability to measure transit depths improves to a precision of 1 ppm, then potential carbon planets around HE\,0107-5240 are the most likely to be spotted (among the group of CEMP-no stars considered in this paper), transiting across the host star at least once every $\sim$ 5 months (10\% condensation efficiency) with a transit duration of 6 days. 

While our calculations place upper bounds on the distance from the host star out to which carbon planets can form,  we note that orbital migration may alter a planet's location in the circumstellar disk. As implied by the existence of `hot Jupiters', it is possible for a protoplanet that forms at radius $r$ to migrate inward either through gravitational interactions with other protoplanets, resonant interactions with planetesimals with more compact orbits, or tidal interactions with gas in the surrounding disk \citep{2006RPPh...69..119P}. 
Since Figure 1 only plots $r_{max}$, the $maximum$ distance out to which a carbon planet with [C/H] can form, our results remain consistent in the case of an inward migration. However, unless planets migrate inward from their place of birth in the disk, we do not expect to find carbon exoplanets orbiting closer than $r \simeq$ 0.02 AU from the host stars since at such close proximities, temperatures are high enough to sublimate carbon dust grains.

Protoplanets can also be gravitationally scattered into wider orbits through interactions with planetesimals in the disk\citep{1999AJ....117.3041H,2009ApJ...696.1600V}. Such an outward migration of carbon planets may result in observations that are inconsistent with the curves in Figure 1. A planet that formed at radius $r \ll r_{max}$ still has room to migrate outwards without violating the `maximum distance' depicted in Figure 1; however, the outward migration of a carbon planet that originally formed at, or near, $r_{max}$ would result in a breach of the upper bounds placed on the transiting properties of carbon planets (Section 5). In particular, a carbon planet that migrates to a semi-major axis $r > r_{max}$ will have an orbital period and a transit duration time that exceeds the limits prescribed in equations (18) and (20), respectively.

Detection of the carbon planets that we suggest may have formed around CEMP stars will provide us with significant clues regarding how planet formation may have started in the early Universe. The formation of planetary systems not only signifies an increasing degree of complexity in the young Universe, but it also carries implications for the development of life at this early junction \citep{2014IJAsB..13..337L}. The lowest metallicity planetary system detected to date is around BD+20 24 57, a K2-giant with [Fe/H] = -1.0 \citep{2009ApJ...707..768N}, a metallicity already well below the critical value once believed to be necessary for planet formation \citep{2001Icar..152..185G,2005MNRAS.364...29P}. More recent formulations of the minimum metallicity required for planet formation are consistent with this observation, estimating that the first Earth-like planets likely formed around stars with metallicities [Fe/H] $\lesssim$ -1.0 \citep{2012ApJ...751...81J}. The CEMP stars considered in this paper are extremely iron-deficient, with [Fe/H] $\lesssim$ -3.2, and yet, given the enhanced carbon abundances which dominate the total metal content in these stars ( [C/H] $\gtrsim$ -1.6), the formation of solid carbon exoplanets in the protoplanetary disks of CEMP stars remains a real possibility. An observational program aimed at searching for carbon planets around these low-mass Population II stars could therefore potentially shed light on the question of how early planets, and subsequently, life could have formed after the Big Bang.

\section{Acknowledgments}
We are thankful to Sean Andrews and Karin \"{O}berg for helpful discussions and feedback. This work was supported in part by NSF grant AST-1312034. This material is based upon work supported by the National Science Foundation Graduate Research Fellowship under Grant No. DGE1144152. Any opinion, findings,
and conclusions or recommendations expressed in this material are those of the authors and do not necessarily reflect the views of the National Science Foundation.


\label{lastpage}
\end{document}